\documentclass[12pt]{article}
\pdfoutput=1
\usepackage{jheppub}

\usepackage{amssymb}
\usepackage{graphicx}
\usepackage{subcaption}
\usepackage{hyperref}
\usepackage{dsfont}
\usepackage{braket}
\usepackage{slashed}
\usepackage{bm}
\usepackage{pifont}
\usepackage{appendix}
\usepackage[table]{xcolor}
\usepackage{multicol}
\usepackage{mathrsfs}

\usepackage{amsmath}
\usepackage{mathrsfs} 
\usepackage{amsfonts}
\usepackage{amssymb}
\usepackage{amsthm}
\usepackage{amsmath}
\usepackage{graphicx}
\usepackage[utf8]{inputenc}
\usepackage[english]{babel}

\usepackage{graphicx}
\usepackage{dcolumn}

\newcommand{\wred}{ {\omega_\infty}}

\newcommand{\shift}{ {\mathcal{D}_{\ssS}} }

\newcommand{\shiftp}{ {\mathcal{D}_{\ssS}^{\prime}} }
\newcommand{\CS}{ {\mathcal{C}_{\ssS}} }
\newcommand{\CSp}{ {\mathcal{C}_{\ssS}^{\prime}} }

\def\smath#1{\text{\scalebox{.85}{$#1$}}}
\def\sfrac#1#2{\smath{\frac{#1}{#2}}}

\usepackage{amsmath}
\usepackage{ulem}
\usepackage{float}

\newcommand{\roughly}[1]{\mathrel{\raise.3ex\hbox{$#1$\kern-0.85em
\lower1ex\hbox{$\sim$}}}}

\newcommand{\gsim}{\roughly>}

\def\ol#1{\overline{#1}}
\def\exd{{\hbox{d}}}

\def\bea{\begin{eqnarray}}
\def\eea{\end{eqnarray}}
\def\be{\begin{equation}}
\def\ee{\end{equation}}

\def\bfu{{\bf u}}

\def\ssB{{\scriptscriptstyle B}}

\def\ssI{{\scriptscriptstyle I}}
\def\ssH{{\scriptscriptstyle H}}

\def\ssM{{\scriptscriptstyle M}}

\def\ssQ{{\scriptscriptstyle Q}}

\def\ssS{{\scriptscriptstyle S}}
\def\ssT{{\scriptscriptstyle T}}
\def\ssU{{\scriptscriptstyle U}}

\def\ssX{{\scriptscriptstyle X}}

\def\\mfK \mfK {{\scriptscriptstyle \mfK \mfK }}

\def\cC{\mathcal{C}}
\def\cD{\mathcal{D}}

\def\cH{\mathcal{H}}
\def\cI{\mathcal{I}}
\def\c\mfK {\mathcal{\mfK }}
\def\cL{\mathcal{L}}

\def\cO{\mathcal{O}}
\def\cP{\mathcal{P}}
\def\cQ{\mathcal{Q}}

\def\cT{\mathcal{T}}

\def\cW{\mathcal{W}}

\def\cZ{\mathcal{Z}}

\def\mfa{\mathfrak{a}}

\def\mfh{\mathfrak{h}}

\def\mfp{\mathfrak{p}}
\def\mfz{\mathfrak{z}}

\def\mfF{\mathfrak{F}}

\def\mfK{\mathfrak{K }}
\def\mfL{\mathfrak{L }}

\def\nn{\nonumber}

\def\({\left(}
\def\){\right)}

\def\pref#1{(\ref{#1})}

\title{Qubits on the Horizon: Decoherence \\and Thermalization near Black Holes}

\author{Greg~Kaplanek}
\author{and C.P.~Burgess,}
\affiliation[a]{Department of Physics \& Astronomy, McMaster University, 1280 Main Street West, Hamilton, Ontario, L8S 4M1, Canada}
\affiliation[b]{Perimeter Institute for Theoretical Physics, 31 Caroline Street North, Waterloo, Ontario, N2L 2Y5, Canada }
\emailAdd{kaplaneg@mcmaster.ca}
\emailAdd{cburgess@perimeterinstitute.ca}
 
\date{}

\abstract{We examine the late-time evolution of a qubit (or Unruh-De Witt detector) that hovers very near to the event horizon of a Schwarzschild black hole, while interacting with a free quantum scalar field. The calculation is carried out perturbatively in the dimensionless qubit/field coupling $g$, but rather than computing the qubit excitation rate due to field interactions (as is often done), we instead use Open EFT techniques to compute the late-time evolution to all orders in $g^2 t/r_s$ (while neglecting order $g^4 t/r_s$ effects) where $r_s = 2GM$ is the Schwarzschild radius. We show that for qubits sufficiently close to the horizon the late-time evolution takes a simple universal form that depends only on the near-horizon geometry, assuming only that the quantum field is prepared in a Hadamard-type state (such as the Hartle-Hawking or Unruh vacua). When the redshifted energy difference, $\omega_\infty$, between the two qubit states (as measured by a distant observer looking at the detector) satisfies $\omega_\infty r_s \ll 1$ this universal evolution becomes Markovian and describes an exponential approach to equilibrium with the Hawking radiation, with the off-diagonal and diagonal components of the qubit density matrix relaxing to equilibrium with different characteristic times, both of order $r_s/g^2$.}

\begin{document}
\maketitle

\section{Introduction and Summary}

Making reliable predictions can be difficult at the best of times. But reliably predicting behaviour at very late times is notoriously hard. What makes it difficult is the inevitable breakdown of perturbative methods that happens at late times; a huge handicap given that perturbative methods dominate a theorist's intellectual toolbag. 

Perturbative methods break down for a simple reason: if a Hamiltonian can be written $H = H_0 + g H_1$ for some small dimensionless parameter $g$, then there is always a time beyond which the time-evolution operator $U(t,t_0) = \exp[-i H(t-t_0)]$ is not well-described by perturbing in $g$. The time where this breakdown occurs scales as an inverse power of $g$, but is eventually exceeded no matter how small $g$ might happen to be. Like death and taxes, perturbative failure is just a matter of time. 

This might not be bothersome if such late times were never of interest. However many important physical processes occur on long time-scales like these. For example, even when individual photons interact weakly with individual atoms, phenomena like refraction and reflection (where 100\% of photons scatter in one direction or another) occur on time-scales long enough to invalidate perturbing in electromagnetic interactions. Thermalization is another phenomenon whose time-scales are very large and scale inversely with coupling strengths like $g$.

In this paper we explore similar late-time issues for interacting quantum systems moving in gravitational fields. That similar phenomena must exist -- particularly in the presence of horizons -- is clear given the thermal nature of quantum fields in these spacetimes \cite{Hawking:1974rv, Hawking:1974sw, Gibbons:1977mu, Israel:1976ur, Sciama:1981hr, Birrell:1982ix}. Test probes should be expected to thermalize in such environments, and any description of this process should share all of the late-time complications that thermalization calculations always have \cite{DaviesOQS, Alicki, Kubo, Gardiner, Weiss, Breuer:2002pc, Rivas, Schaller}. In this paper we show this is true for quantum systems exterior to a Schwarzschild black hole, extending our own earlier work that does so for spacetimes with Rindler \cite{Burgess:2018sou, Kaplanek:2019dqu} and de Sitter horizons \cite{Kaplanek:2019vzj}.

The reason for doing so is not because this kind of thermalization is soon likely to be observed. On the contrary, it is worth doing because the tools used are informative in their own right. In particular, they show how standard techniques used to describe late-time behaviour in optics and thermal physics apply equally well in gravitational settings \cite{Starobinsky:1986fx, Salopek:1990re, Starobinsky:1994bd, Polarski:1995jg, Burgess:2009bs, Burgess:2010dd, Burgess:2014eoa, Agon:2014uxa, Burgess:2015ajz, Agon:2017oia, TheBook}. This makes them potentially relevant to late-time puzzles known to occur in gravity, such as the problem of secular growth in cosmological spacetimes \cite{Ford:1985, Ford:1985qh, Antoniadis:1986, Muller:1988, Antoniadis:1991, Sasaki:1993, Dolgov:1995, AB, Wbg, Sloth, Bilandzic:2007nb, Lyth:2007jh, BMPRS, RS, Enqvist:2008kt, LU, JMPW, Adshead:2008gk, UraTan, DGLAP, Senatore:2009cf, Giddings:2010nc, Byrnes:2010yc, Gerstenlauer:2011ti} (for reviews see  \cite{Seery:2010kh, Tanaka:2013caa}) and to problems like information loss \cite{Hawking:1976ra} or `firewall' problems \cite{Almheiri:2012rt, Almheiri:2013hfa} in black-hole physics (for reviews see \cite{Banks:1994ph, Mathur:2009hf}).

In this paper we compute the late-time evolution of a two-level quantum system ({\it i.e.}~a qubit or Unruh detector \cite{Unruh:1976db, DeWitt:1980hx, Sciama:1981hr}) that hovers at fixed radius $r = r_0$ above the event horizon of a Schwarzschild black hole while interacting with a quantum scalar field. We do so perturbatively in the dimensionless coupling strength $g$ with which the qubit interacts with the quantum field. We show that if $\wred$ is the redshifted splitting of the two qubit energy levels, and if $\wred \,r_s \ll 1$ where $r_s = 2GM$ is the usual Schwarzschild radius, and if the scalar-field mass also satisfies $m \,r_s \ll 1$, then such a qubit has universal late-time behaviour (for $t \gsim r_s/g^2$) provided that it sits sufficiently close to the event horizon: $0 < r_0 - r_s \ll r_s$. 

Not surprisingly, this universal evolution describes the evolution of the qubit towards an asymptotic thermal state whose temperature equals the Hawking temperature $T = T_\ssH := (4\pi r_s)^{-1}$. Perhaps more surprisingly we show that this approach to equilibrium is also very robust, occurring exponentially with two different thermalization time-scales proportional to
\begin{eqnarray} \label{xiresult}
\xi  =   \frac{4\pi \tanh\left( 2 \pi r_s\, \wred \right) }{g^2 \wred}   \simeq   \frac{8\pi^2 r_s}{ g^2  } + \cdots \quad \hbox{since $\wred r_s \ll 1$}  \,.
\end{eqnarray}
This evolution is robust in the sense that it depends only on the qubit/field coupling strength, $g$, and on the background geometry for any quantum state whose Wightman function has the standard `Hadamard' form \cite{Hadamard, DeWitt:1960fc, Fulling:1978ht} at small field separations: {\it i.e.} it satisfies \pref{hadamard}, reproduced here as
\be \label{hadamardprev}
 G_\Omega (x,x') =  \frac{ 1 }{8 \pi^2 \,  \sigma(x,x') } + \cdots  \,,
\ee
where $\sigma(x,x')$ is half the square of the geodesic distance between spacetime points $x$ and $x'$. In particular, eq.~\pref{xiresult} applies equally well if the quantum field is prepared in either the Hartle-Hawking or Unruh vacua, and is independent of the scalar-field mass in the mass range $m\, r_s \ll 1$. 

It has been known for some time that Hadamard behaviour suffices for deriving the steady-state Hawking flux around Schwarzschild black holes \cite{Fredenhagen:1989kr}, and our results extend this conclusion to the approach to equilibrium for quantum probes. We remark in passing that our results differ from early -- and some recent -- calculations of Unruh detectors in Schwarzschild geometries \cite{Candelas:1980zt, Hodgkinson:2012mr, Ng:2014kha, Ng:2017iqh, Emelyanov:2018woe, Jonsson:2020npo}, which often compute qubit excitation rates, finding results that differ when the field is prepared in different states (such as the Hartle-Hawking or Unruh vacua). These calculations usually compute the rate with which a qubit is excited out of its ground state, as opposed to the qubit's late-time approach to its asymptotic thermal state (as is computed here). Although the excitation rate can be accessed perturbatively in $g$, more effort is required to obtain the approach to equilibrium since the time-scale involved is of order $r_s/g^2$. 

We are able to make reliable predictions using arguments of Open Effective Field Theories (Open EFTs) \cite{Burgess:2014eoa, Burgess:2015ajz, TheBook}. As is explained in more detail in \cite{Kaplanek:2019dqu}, these recast techniques from elsewhere in physics into an effective field theory language that is easily adapted to gravitational systems. In essence these arguments have a renormalization-group like structure: one sets up a differential evolution equation for the object of interest (in this case the reduced density matrix for the qubit) whose domain of validity is larger than the integrated evolution from which it is derived. That is, one explicitly evolves the system using perturbation theory starting from an arbitrary initial time, $t_0$. Although perturbative evolution can only be used to evolve a limited way forward in time, say from $t_0$ to $t_1$, within this window the result can be differentiated with respect to time to derive a differential evolution equation. 

If this evolution equation itself makes no specific reference to $t_0$ then the same construction could equally well be used to derive the same evolution equation starting at $t_1$, with perturbative validity out to $t_2$, and again starting at $t_2$ and so on. Whenever this can be done the solutions to the differential evolution equation can be valid on the union of each of these derivation intervals. If $g \ll 1$ is the small perturbative expansion parameter then this process ends up resumming all orders in $g^2 t$, say, but neglecting contributions in the evolution\footnote{Order $g^4 t$ evolution is similarly predicted using a more accurate evolution equation, and so on.} that are of order $g^4 t$. As a result the solutions found this way can be trusted even when $t \sim \cO(r_s/g^2)$.

One reason to explore the simple qubit systems considered here is to make this construction very explicit, making it easier to understand. The starting point for the argument is  the Nakijima-Zwanzig equation \cite{Nak, Zwan}, which is a general evolution equation for the reduced $2 \times 2$ density matrix, $\boldsymbol{\varrho}(t)$, of the qubit. It is obtained by tracing over the Liouville equation describing the evolution of the full qubit/field system, and then eliminating that part of the density matrix that describes the non-qubit degrees of freedom. The result is an integro-differential evolution equation that is useful because it refers only to the qubit's reduced density matrix and not to the other degrees of freedom, which appear only implicitly through correlation functions of $H_{\rm int}$. Although the Nakajima-Zwanzig equation does not in itself automatically allow perturbative time-evolution to be extended out to very late times, it provides a useful starting point for identifying situations where this can be done.

As is true for most effective field theories, relative s implicity comes only when there is a hierarchy of scales that can be exploited. The important hierarchy arises in this case if the field correlation function $\langle H_{\rm int}(t) H_{\rm int}(t') \rangle$ falls off to zero for $|t-t'| > \zeta$, for some characteristic time-scale $\zeta$. In this case the useful hierarchy arises when exploring time-evolution over much longer time-scales $\Delta t \gg \zeta$. Access to late times can happen if the Nakajima-Zwanzig equation remains sufficiently simple once expanded in powers of this ratio $\zeta/\Delta t$. 

The qubit example studied here shows in detail how this can happen: the leading terms in the Nakajima-Zwanzig equation become Markovian, in the sense that $\partial_t \boldsymbol{\varrho}(t)$ depends only on $\boldsymbol{\varrho}(t)$ and not on the details of its past history prior to time $t$. Markovian behaviour of this form emerges for qubits near a black hole once $\Delta t \gg r_s$ (at least this is true when the redshifted energy difference $\omega_\infty$ between the two qubit energy levels -- as seen by a static observer looking at the qubit far from the black hole -- satisfies $\omega_\infty r_s \ll 1$), Evolution to all orders in $g^2 t$ is then described by a Lindblad equation \cite{Lindblad:1975ef, Gorini:1976cm}. (Some implications of Lindblad evolution in Schwarzschild geometries are also explored in  \cite{Yu:2008zza, Chatterjee:2019kxg, Singha:2018vaj, Feng:2015xza, Hu:2012gv, Zhang:2011vsa, Hu:2011pd}.) By deriving the Lindblad equation as a limit of the Nakajima-Zwanzig equation for this system, we are able to assess its domain of validity. 

\subsubsection*{This paper}

The rest of this paper is organized as follows. The next section, \S\ref{sec:QiS}, sets up the system whose late-time near-horizon evolution is to be computed. In particular \S\ref{sec:QiS} defines our qubit/quantum-field system for static spacetimes, and then briefly explores the properties of qubit trajectories that hover at fixed positions just above a Schwarzschild black hole. 

\S\ref{sec:OpenSys} follows this with a brief description of how reduced density matrices are evolved in open systems, describing the Nakajima-Zwanzig equation whose solutions govern the qubit's late-time behaviour. Since at lowest nontrivial order the quantum field enters into the qubit evolution only through its Wightman function, we also summarize in \S\ref{sec:OpenSys} the near-horizon form for this function for field states that satisfy the Hadamard form for small separations. 

Finally, \S\ref{sec:Universal} shows how the near-horizon limit of the Wightman function allows the Nakajima-Zwanzig equation to be approximated by a Markov process, describing the late-time exponential decay towards a Hawking-temperature thermal state. The time-scale for this approach to equilibrium is computed for qubits asymptotically close to the horizon, and found to be universal in the sense that it is determined only by qubit properties and the black-hole geometry. Provided $m \, r_s \ll 1$ this rate is largely independent of the details of the quantum field, and assumes only that it is prepared in a Hadamard state. In particular the approach to equilibrium is the same when the field is prepared in either a Hartle-Hawking or Unruh state. 

\section{Qubits in Schwarzschild}
\label{sec:QiS}

This section sets up the framework -- a qubit/field system and the spacetime through which the qubit moves -- that is used to perform the calculations to follow.

\subsection{The qubit/scalar system}

The system whose evolution we follow consists of a real massive scalar field $\phi(x)$ coupled to a single two-level qubit through the action $S = S_\ssB + S_\ssQ + S_{\rm int}$, where $S_\ssB$ describes a self-interacting quantum field
\be \label{SBdef}
  S_\ssB =  - \frac{1}{2} \int \exd^4 x\; \sqrt{ - g} \bigg[ g^{\mu\nu} \partial_{\mu} \phi \, \partial_{\nu} \phi +  ( m^2 + \xi R ) \phi^2 + \frac\lambda{12} \, \phi^4 \bigg]  \,,
\ee
within a background metric\footnote{Nothing precludes also quantizing the fluctuations of the metric about the given background, using standard EFT arguments \cite{Weinberg:1978kz, Donoghue:1994dn, Burgess:2003jk, Donoghue:2017ovt}, though for simplicity we do not do so since we do not expect this not to alter our main point. In principle, dropping metric fluctuations can justified quantitatively by working with $N \gg 1$ scalar fields and computing in the leading large-$N$ limit.} given as in \pref{metricS} or \pref{metricK}. For this paper we neglect self-interactions ($\lambda = 0$), though we briefly comment in the conclusions on how things can change in their presence. The coupling $\xi$ plays no role because Schwarzschild is Ricci flat, and for reasons to be clear below the mass $m$ is assumed to satisfy $m \, r_s \ll 1$.

The free qubit action is given by\footnote{Terms linear in $\mfz_i$ do not appear in this action because we require the classical Grassmann action to be Grassman-even.} 
\be
  S_\ssQ = \int \exd^4 x \int \exd s \left[\frac{i}2 \,\cZ\, \mfz^i \, \dot \mfz_i   - \sqrt{ - \dot y^2} \left( \omega_0 - \frac{i \hat\omega}4 \, \epsilon^{ijk} u_i \, \mfz_j \mfz_k \right)\right] \delta^4[x - y(s)] \,,
\ee
where $\mfz_i(s)$ are classical Grassman variables (with $i = 1,2,3$) with $\dot \mfz_i := \exd \mfz_i/\exd s$ and $\mfz^i := \delta^{ij} \mfz_j$. The quantities $\cZ$, $\omega_0$ and $\hat\omega$ and $u^i$ are real parameters (with $u_i u^i = 1$), with $\cZ$ eventually absorbed into the $\mfz_i$ to obtain a convenient normalization that simplifies later formulae.\footnote{Hats appear on couplings like $\hat \omega$ and $\hat g$ to distinguish them from the corresponding quantities once appropriate powers of $\cZ$ have been absorbed into $\mfz_i$.} The integral over $\exd^4x$ is trivially done using the delta-function, and reveals that the integration is over a specific timelike world line $x^\mu = y^\mu(s)$, along which the qubit moves through the ambient spacetime.\footnote{For real systems $y^\mu(s)$ is itself a dynamical variable to be quantized, but for simplicity we ignore this complication here (treating the qubit trajectory as being specified), since it does not affect our later discussions. } Here $s$ is a  parameter along this world line, and the quantity 
\be
   \dot y^2 := g_{\mu\nu}[y(s)] \, \dot y^\mu \,\dot y^\nu 
\ee
is what is required to ensure that $S_\ssQ$ is invariant under reparameterizations of $s$. It is usually convenient to fix this freedom by choosing proper time, $\tau$, along the curve as the parameter, in which case $\dot y^2 = -1$. 

Interactions beween $\mfz$ and $\phi$ are assumed to take the form
\be \label{Sintdef}
   S_{\rm int}  = \frac{i\hat g}2 \int \exd^4x \int \exd \tau \; \phi \, \epsilon^{ijk} n_i \mfz_j \mfz_k \;  \delta^4[x-y(\tau)]= \frac{i\hat g}2  \int \exd \tau \; \phi[y(\tau)] \, \epsilon^{ijk} n_i \mfz_j (\tau)\mfz_k (\tau) \,,
\ee
where $\hat g$ and $n^i$ are real coupling constants, with $n_i n^i = 1$, and our analysis is ultimately performed perturbatively in $\hat g$.

\subsubsection*{Quantization}

Working in the interaction picture, quantization of $\mfz_i$ and $\phi$ is performed as if they did not interact, with interactions included in powers of $g$ (and $\lambda$) once time-evolution is evaluated. 

For $\mfz_i$ this quantization goes through as usual \cite{Zalavari:2020tez}, keeping in mind this is a constrained system. To see why, recall that the canonical momenta are given by
\be \label{cononpvsz}
  \mfp^i := \frac{\delta S_\ssQ}{\delta \dot \mfz_i} = \frac{i \cZ}{2} \;  \mfz^i  \,.
\ee
Because this cannot be solved for $\dot \mfz_i$ as a function of the $\mfp^j$ it is instead regarded as a constraint: $\mfp^i - \frac{i}2 \cZ\, \mfz^i = 0$. The qubit hamiltonian (generating evolution in proper time $\tau$) becomes
\be \label{mfhdef}
  \mfh = \mfp^i \dot \mfz_i - \left[\frac{i}2 \, \mfz^i \, \dot \mfz_i   - \left( \omega_0  - \frac{i\hat\omega}4 \epsilon^{ijk} u_i \, \mfz_j \mfz_k \right)\right] = \omega_0 - \frac{i\hat\omega}4 \epsilon^{ijk} u_i \, \mfz_j \mfz_k \,.
\ee
The canonical quantization conditions turn out to imply that the anticommutator of $\mfp^i$ and $\mfz_j$ is proportional to $\delta^i_j$, and the parameter $\cZ$ can be chosen to ensure that the quantum version of the Grassmann condition becomes
\be \label{PauliAlgebra}
  \Bigl\{ \mfz_i \,, \mfz_j \Bigr\} = 2 \, \delta_{ij} \,.
\ee

The space of quantum states furnishes a representation of this algebra, and for a 2-level qubit this representation is two-dimensional. The required operator representations for the $\mfz_i$ therefore are the Pauli matrices
\be
   \mfz_1 = \boldsymbol{\sigma}_1 = \left( \begin{matrix} 0 & 1 \\ 1 & 0 \end{matrix} \right) \,, \quad 
   \mfz_2 = \boldsymbol{\sigma}_2 = \left( \begin{matrix} 0 & -i \\ i & 0 \end{matrix} \right)  \quad \hbox{and}\quad
   \mfz_3 = \boldsymbol{\sigma}_3 = \left( \begin{matrix} 1 & 0 \\ 0 & -1 \end{matrix} \right)  \,.
\ee
With this choice, properties of the Pauli matrices ensure that $\epsilon^{ijk} \mfz_j \mfz_k = 2i \mfz^i$, and so defining coordinates so that $\bfu \cdot \boldsymbol{\sigma} = \boldsymbol{\sigma}_3$ then allows \pref{mfhdef} to be written explicitly as
\be \label{mfheval}
  \mfh = \omega_0 \, \mathbb{I}  + \frac{\omega}2  \, \boldsymbol{\sigma}_3 \,,
\ee
where $\mathbb{I}$ is the $2\times 2$ unit matrix, and hats are dropped on $\omega$ when variables are normalized so that \pref{PauliAlgebra} holds. Eq.~\pref{mfheval} reveals the free-qubit energy eigenvalues to be $\omega_0 \pm \frac12 \, \omega$ and so $\omega_0$ is their mean energy while $\omega$ (which we take to be positive) gives their level splitting (as measured by an observer whose time is the qubit's proper time, $\tau$). 

With this representation for $\mfz_i$ the qubit/field interaction \pref{Sintdef} becomes
\be \label{Sinteval}
   S_{\rm int}  = -g \int \exd \tau \; \phi[y(\tau)] \, n^i \boldsymbol{\sigma}_i = -g \int \exd \tau \; \phi[y(\tau)] \, \boldsymbol{\sigma}_1\,,
\ee
where the second equality specializes to the case where $n^i$ is perpendicular to $u^i$ (and so can be chosen to lie along the `$1$' axis). 

In the absence of couplings ($\lambda = g = 0$) the scalar field is quantized in the usual fashion for a static curved space \cite{Birrell:1982ix}. The interaction-picture field equation is
\be \label{KGeq}
\big(- \Box + m^2 + \xi R \big) \phi = 0  \,,
\ee
and so for a static geometry one expands 
\be
  \phi(x) = \sum_n \Bigl[ u_n(x) \, \mfa_n + u_n^*(x) \, \mfa_n^* \Bigr] \,,
\ee
where $u_n$ simultaneously satisfies $\mfL_t u_n = -i \omega_n u_n$ and eq.~\pref{KGeq}, where $\mfL_t$ is the symmetry generator in the timelike direction along which the metric is static. Canonical commutation relations imply the creation and annihilation operators satisfy $[ \mfa_n \,, \mfa^*_m ] = \delta_{mn}$.  As mentioned earlier, for applications to Schwarzschild our interest is in small masses,\footnote{Physically, once the scalar mass becomes much bigger than the temperature its states become exponentially rarely occupied, leading one to expect them to decouple from qubit evolution. This expectation can be very explicitly verified for simple systems such as an accelerating qubit moving through flat spacetime \cite{Kaplanek:2019dqu}.} $m\, r_s \ll 1$, and Ricci-flatness makes the term $\xi R$ drop out of subsequent discussion.

The scalar field hamiltonian (including self-interactions) is easily computed in the presence of any spacetime metric of the form
\be \label{staticmetric}
   \exd s^2 = - f \, \exd t^2 + \gamma_{ab} \, \exd x^a \, \exd x^b 
\ee
where $g_{at} = 0$ and both $g_{tt} = -f$ and the spatial metric $g_{ab} = \gamma_{ab}$ are $t$-independent. The hamiltonian density (for the generator of evolution in $t$) is then given by
\be
  \cH = \Pi \partial_t \phi - \cL 
\ee
where $\cL$ is the lagrangian density from \pref{SBdef} and the canonical momentum is
\be 
   \Pi := \frac{\delta S_\ssB}{\delta \partial_t \phi} = \sqrt{\frac{\gamma}{f}} \; \partial_t \phi  \,,
\ee
where \pref{SBdef} is again used, together with the metric \pref{staticmetric}, to evaluate the derivative. Therefore the scalar-field hamiltonian density, $\cH$, becomes
\be \label{HBdef}
  \cH = \sqrt{f \gamma} \; \left[ \frac{(\partial_t \phi)^2}{2f} + \frac{1}{2} g^{ab} \partial_a \phi \, \partial_b \phi + \frac12(m^2 + \xi R) \phi^2 + \frac{\lambda}{4!} \phi^4 \right] \,.
\ee

\subsubsection*{Total energy}

We can now assemble everything to identify the total field/qubit hamiltonian, leading to the following sum:
\be
  H = H_0 + H_{\rm int}
\ee
where the `free' hamiltonian is the sum of the two free hamiltonians constructed above
\be \label{H0def}
H_0  = \cH \otimes \mathbb{I} + \cI \otimes \mfh \;\frac{\exd \tau}{ \exd t} \,.
\ee
Here $\cI$ is the unit operator in the scalar-field state-space, while $\cH$ and $\mfh$ are as given in eqs.~\pref{HBdef} and \pref{mfheval}. The factor $\exd \tau/\exd t$ is required because $\mfh$ generates translations in proper time $\tau$ while $H_0$ (and $\cH$) are defined to generate translations in $t$. 

Since the interaction Lagrangian does not involve time derivatives its contribution to the hamiltonian is simple to write down, starting from \pref{Sinteval}
\be \label{Hinteval}
   H_{\rm int}  = g \; \phi[y(\tau)] \otimes n^i \boldsymbol{\sigma}_i \;\frac{\exd \tau}{ \exd t} =  g \;  \phi[y(\tau)] \otimes \boldsymbol{\sigma}_1 \;\frac{\exd \tau}{ \exd t}\,.
\ee
In what follows we compute the implications of this interaction out to second order in $g$. The nature of this perturbation theory depends on the relative size of $\omega$ and the $\cO(g^2)$ corrections to the qubit energy levels, and for simplicity we work in the regime where these corrections are much smaller than the qubit's zeroth-order level splitting, a restriction that eventually leads to the parameter conditions summarized in Table \ref{MarkovianTable0}.

\subsection{Near-horizon geometry}

Our interest lies in the near-horizon limit of the exterior of a spinless black hole, defined in Schwarzschild coordinates by
\be 
\exd s^2  =  - \left( 1- \frac{r_{s}}{r} \right) \exd t^2 + \left( 1- \frac{r_{s}}{r} \right)^{-1} \exd r^2 + r^2 \exd \theta^2 + r^2 \sin^2 \theta \, \exd \phi^2 \label{metricS} \,,
\ee
where $r > r_s$, where $r = r_s  := 2GM$ defines the event horizon. These coordinates are useful because they fall into the category defined in \pref{staticmetric}, with $t$ being the static time on which the metric does not depend, and so $\mfL_t = \partial_t$. The (Kretchsmann) curvature invariant for this geometry is $R_{\rho\sigma\tau\kappa} R^{\rho \sigma \tau \kappa} = {12r_{s}^2}/{r^6}$, and so is nonsingular at $r = r_s$.

Schwarzschild coordinates famously break down at the horizon, in the vicinity of which Kruskal coordinates, $(T,X,\theta,\phi)$, defined by
\bea \label{KruskalTX}
T &=&  \sqrt{ \frac{r}{r_{s}} - 1 } \;  \exp\left( \frac{r}{2r_{s}} \right) \sinh\left( \frac{t}{2 r_{s}} \right) \nn\\
 X &=&  \sqrt{ \frac{r}{r_{s}} - 1 } \; \exp\left( \frac{r}{2r_{s}} \right) \cosh\left( \frac{t}{2 r_{s}} \right) \,,
\eea
are more useful. In terms of these the line element becomes
\be \label{metricK}
\exd s^2 = \frac{4 r_{s}^3}{r} \; e^{- {r}/{r_{s}}} \left( - \exd T^2 + \exd X^2 \right) + r^2 \exd \theta^2 + r^2 \sin^2 \theta \, \exd \phi^2 
\ee
where now $r = r(X,T)$ is the implicit function of $X$ and $T$ given by solving 
\be \label{r0inKrusk}
  X^2 - T^2 = \left( \frac{r}{r_{s}} - 1 \right) e^{ {r}/{r_{s}}}   \,,
\ee
and so is given by 
\be  
  r(X,T)  = r_s \Bigl[ 1 + \cW(z) \Bigr] \quad \hbox{where} \quad z := \frac{1}{e} (X^2 - T^2) \,,
\ee
and $\cW$ is the Lambert $W$ function defined by $\cW(z) \exp[\cW(z)] = z$ (which has a unique real solution for $z > 0$ and two real branches for $-e^{-1} < z < 0$).  

Although these coordinates are well-behaved at the horizon, the spatial geometry at fixed $T$ is $T$-dependent. This is a reflection of the coordinate-independent statement that the metric is only static outside of the horizon (where it can be rewritten as \pref{metricS}). Inside the horizon the metric's symmetry directions are all spacelike.

\subsubsection*{Hovering world-lines}

The qubits whose late-time evolution we follow are chosen to hover at fixed $r = r_0 > r_s$ above the event horizon. This is clearly not a geodesic and so the qubit can be maintained along this trajectory through the action of some non-gravitational force, whose detailed nature need not concern us here. 

Consider two events on this world-line that are distinguished by two values, $\tau_1$ and $\tau_2$, of the qubit's proper time; separated by $\Delta \tau := \tau_2 - \tau_1 >0$. These two events are separated by a redshifted time $\Delta t$ for hovering observers situated at spatial infinity, with \pref{metricS} implying that 
\be 
   \Delta \tau :=  \int_{t_1}^{t_2} \exd t \; \sqrt{ - g_{\mu\nu} \frac{\exd y_0^\mu}{\exd t} \frac{\exd \dot y_0^\nu}{\exd t}}  = \Delta t \sqrt{ 1 - \frac{r_s}{r_0} } \,, \label{timerelation}
\ee 
where $y_0^\mu(t)$ denotes the curve along which only $t$ varies, with $r$, $\theta$ and $\phi$ all fixed. Notice, for future use, that $\Delta \tau \ll \Delta t$ when $0 < r_0 - r_s \ll r_s$. 

Both of these intervals differ from the geodesic separation of these two points, 
\be
   \Delta s :=  \int_{t_1}^{t_2} \exd t \; \sqrt{ - g_{\mu\nu} \frac{\exd y_g^\mu}{\exd t} \frac{\exd \dot y_g^\nu}{\exd t}} \,,
\ee
where the subscript `$g$' indicates that integration is evaluated along the geodesic $y_g^\mu(t)$ that satisfies
\be
  \ddot y_g^\mu + \Gamma^\mu_{\nu\lambda} \dot y_g^\nu \dot y_g^\lambda = 0 
\ee
as well as $\theta(t) = \theta_0$, $\phi(t) = \phi_0$ for all $t$ while $r(t_1) = r(t_2) = r_0$. Because this is a geodesic it must describe the longest time interval as measured along any timelike curve that connects the two events, so $\Delta s > \Delta \tau$. Such a geodesic is possible if $\exd r/\exd t(t_1) > 0$ is chosen appropriately, since then any freely falling body initially moves radially away from the horizon before eventually turning back and falling into the black hole. 

In what follows it proves more convenient to work with the Synge world function, $\sigma(x_1, x_2)$, defined for timelike geodesics by \cite{J.L.Synge:1960zz, Poisson:2011nh, Ottewill:2008uu}
\be
  \sigma(x_1, x_2) = - \frac12 \, (\Delta s)^2 \,,
\ee
since this has an integral form that is easier to manipulate (see Appendix \ref{App:Synge}). 

For later purposes we are interested in a limit that simultaneously has a small invariant interval, $\Delta s \ll r_s$, but corresponds to late times $\Delta t \gg r_s$. We remark in passing that the above formulae show that both of these can be simultaneously true provided we pick $r_0 > r_{s}$ sufficiently close to the horizon so that
\be \label{close}
1 - \frac{r_{s}}{r_0}  \ll  1 \,.
\ee
Being close to the horizon suffices because  the curve that hovers at fixed $r = r_s$ {\it is} a geodesic, although it is a null geodesic -- for which $\Delta s = 0$ -- rather than a timelike one.  

\section{Time evolution in open systems}
\label{sec:OpenSys}

We return now to the evolution of the qubit that hovers just above the horizon while interacting with the quantum field. Our interest is in how this qubit responds to the fluctuations of the quantum field, and in how this response becomes universal in the late-time limit very near the horizon. 

Since it is only the qubit's behaviour that is to be predicted, it is convenient to trace out the scalar field from the system density matrix, and work instead only with the qubit's $2\times 2$ reduced density matrix, defined as
\begin{eqnarray}
\boldsymbol{\varrho}(t) & : = & \underset{\phi}{\mathrm{Tr}} \left[ \rho(t) \right]
\end{eqnarray}
where $\rho(t)$ is the total -- {\it i.e.}~the combined field/qubit -- density matrix, and the trace is over the scalar-field part of the Hilbert space. 

When needed we assume the field and qubit to be initially uncorrelated, 
\be 
\rho_0 := \rho(t_{\rm i}) = \Omega \otimes \boldsymbol{\varrho_0}
\ee
where $\boldsymbol{\varrho_0}$ defines the initial qubit state and $\Omega$ is the density matrix for the quantum field. Three commonly made choices for $\Omega$ might be the Hartle-Hawking state, $\Omega_\ssH := | \mathbb{H} \rangle \langle \mathbb{H} |$, the Unruh state $\Omega_\ssU := | \mathbb{U} \rangle \langle \mathbb{U} |$ or the Boulware state, $\Omega_\ssB := | \mathbb{B} \rangle \langle \mathbb{B} |$. These are all pure states that are candidate vacua for the field, with $|\mathbb{H} \rangle$ corresponding to the vacuum in the presence of a black hole that is in equilibrium with a bath of radiation prepared at the Hawking temperature, while $|\mathbb{U} \rangle$ is the late-time vacuum for a black hole that forms in isolation. 

Time evolution for $\boldsymbol{\varrho}$ is in principle determined by the evolution of the full system's density matrix, which in the interaction picture satisfies
\be \label{LiouvilleDE}
   \partial_t \rho(t) = -i \Bigl[ V(t) \,, \rho(t) \Bigr] \,,
\ee
where $V(t) := e^{i H_0 t} H_{\rm int} e^{-i H_0 t}$. Eq.~\pref{LiouvilleDE} has a standard perturbative solution
\be \label{Liouville}
  \rho(t) = \rho_{0} -i \int_{t_{\rm i}}^t \exd s \Bigl[ V(s) \, \rho_{0} \Bigr] - \frac12 \int_{t_{\rm i}}^t \exd s_1 \int_{t_{\rm i}}^{s_1} \exd s_2 \; \Bigl[ V(s_2) \,, \Bigl[ V(s_1) \, \rho_{0} \Bigr] \Bigr] + \cdots \,,
\ee
given the initial condition $\rho(t_{\rm i}) = \rho_0$. 

As is discussed at great length elsewhere -- see for example \cite{Kaplanek:2019dqu, Kaplanek:2019vzj, TheBook} and references therein -- there are two major obstacles to using eqs.~\pref{LiouvilleDE} or \pref{Liouville} to predict $\boldsymbol{\varrho}(\tau)$ at late times. 
\begin{enumerate}
\item At first sight one could trace \pref{LiouvilleDE} over the scalar-field sector to obtain $\partial_t \boldsymbol{\varrho}$, but the result is hard to solve for $\boldsymbol{\varrho}(\tau)$, because the dependence of the right-hand side on $\boldsymbol{\varrho}$ is only implicitly given through its dependence on the full density matrix $\rho$.
\item The solution \pref{Liouville} does not have this same difficulty because in his equation we may use $\rho_0 = \Omega \otimes \boldsymbol{\varrho_0}$. The problem with \pref{Liouville} is that the perturbative approximation on which it relies systematically breaks down at very late times (in the present example this breakdown occurs at times of order $t \sim r_s/g^2$). 
 \end{enumerate}

The Nakajima-Zwanzig equation \cite{Nak, Zwan} provides the solution to problem (1) above, and this is useful because the result shows how to find solutions that are not afflicted by problem (2), in that they allow reliable perturbative predictions even when $t$ is so large that $g^2 t$ cannot be neglected relative to $r_s$.
   
\subsection{The Nakajima-Zwanzig Equation}

The logic of the Nakajima-Zwanzig equation is to project the full density matrix onto the reduced density matrix and its complement: %
\be
   \boldsymbol{\varrho} = \cP(\rho) \quad \hbox{and} \quad
   \Xi := \cQ(\rho) 
\ee
for some projection operator $\cP^2 = \cP$ and the second definition uses $\cQ := 1 - \cP = \cQ^2$. Since the time-evolution equation  \pref{LiouvilleDE} for $\rho$ is a linear equation it can be turned into a pair of coupled linear evolution equations for the two quantities $\boldsymbol{\varrho}$ and $\Xi$. Eliminating $\Xi$ from this system gives the Nakajima-Zwanzig equation: an evolution equation that involves only $\boldsymbol{\varrho}$, but  is nonlocal in time due to the elimination of $\Xi$. Because this is essentially a linear problem, it can be solved in great generality \cite{Nak, Zwan}.

As applied to the current example, following identical steps as given in \cite{Kaplanek:2019dqu, Kaplanek:2019vzj} leads to the following result at second order in the coupling $g$:
\begin{eqnarray} \label{INTpictureNZ} 
\frac{\partial \boldsymbol{\varrho}^\ssI (\tau)}{\partial \tau} &\simeq& g^2 \int_0^\tau \exd s\ \bigg( G_\Omega(\tau,s)\ \big[ \mathfrak{m}^\ssI(s) \, \boldsymbol{\varrho}^\ssI (s), \mathfrak{m}^\ssI(\tau) \big]  \\
 && \qquad\qquad\qquad\qquad  +G^{\ast}_\Omega(\tau, s) \big[ \mathfrak{m}^\ssI(\tau) , \boldsymbol{\varrho}^\ssI (s)\, \mathfrak{m}^\ssI(s)  \big] \bigg) - i \left[ \frac{\delta \omega}{2} \boldsymbol{\sigma_3} , \boldsymbol{\varrho}^{\ssI}(\tau) \right] \, , \nn
\end{eqnarray}
where $\boldsymbol{\varrho}^{\ssI}(\tau) = e^{+ i \mathfrak{h} \tau} \boldsymbol{\varrho}(\tau) \,e^{- i \mathfrak{h} \tau}$ is the reduced density matrix in the interaction-picture representation and so similarly $\mathfrak{m}^{\ssI}(\tau) := e^{+ i \mathfrak{h} \tau} \boldsymbol{\sigma_1} e^{- i \mathfrak{h} \tau}$ (and conventions generally follow \cite{Kaplanek:2019dqu,Kaplanek:2019vzj}). 
At this point several features of \pref{INTpictureNZ} bear explanation.
\begin{itemize}
\item First, notice that eq.~\pref{INTpictureNZ} gives the evolution of $\boldsymbol{\varrho}$ as a function of proper time along the qubit trajectory, and does so despite its derivation starting from the Liouville equation \pref{LiouvilleDE}, which is phrased in terms of the geometry's static time coordinate, $t$. This occurs in detail because of the time-dilation factors $\exd \tau/\exd t$ that appear in the Hamiltonian in eqs.~\pref{H0def} and \pref{Hinteval}.
\item Second, in expression \pref{INTpictureNZ} the quantity $G_\Omega(\tau, s)$ represents the scalar-field Wightman functions 
\be
  G_\Omega(\tau, s) :=  \underset{\phi}{\mathrm{Tr}} \Bigl( \Omega \;  \phi[ y(\tau) ] \phi[ y(s) ] \Bigr) \,,
\ee
evaluated at two places along the qubit trajectory, $y^\mu(s)$, and we use the property $G_\Omega(s,\tau) = G^{\ast}_\Omega(\tau, s)$ for the Wightman function of a real scalars. These are fairly complicated functions when evaluated in Schwarzschild spacetime and are usually given implicitly in terms of a sum over mode functions \cite{Candelas:1980zt, Page:1982fm, Candelas:1984pg, Matyjasek:1998mq, Boulware:1974dm, DeWitt:1975ys, Hartle:1976tp, Unruh:1976db, Christensen:1977jc, Yu:2008zza}. In what follows we choose a near-horizon trajectory along which they take a simple form. 
\item Third, the final term in \pref{INTpictureNZ} comes from a counter-term interaction, obtained by replacing $\omega \to \omega_{\rm bare} = \omega + \delta \omega$ in $\mfh$, with $\delta \omega$ regarded as being $\cO(g^2)$. This counter-term is required because the qubit/field interaction shifts the inter-level energy spacing, and so makes the parameter $\omega$ appearing in $\mfh$ no longer equal to this spacing. If the parameter in $\mfh$ is therefore instead called $\omega_{\rm bare}$ then $\omega$ remains the physical spacing of qubit levels if $\delta \omega$ is chosen to cancel the order $g^2$ qubit energy shift.\footnote{A bonus of this definition is that $\delta \omega$ also automatically cancels an ultraviolet divergence that arises in the computed energy shift.}
\item Finally, notice that \pref{INTpictureNZ} would agree with the time derivative of \pref{Liouville} if in the right-hand side one were to replace $\varrho^\ssI(s)$ with its initial condition $\varrho^\ssI_0$. Furthermore, such a replacement at face value seems to be compulsory, because the difference between $\varrho^\ssI(s)$ and $\varrho^\ssI_0$ is higher order in $g$. It is this assumption that $\varrho^\ssI(s)$ and $\varrho^\ssI_0$ are only perturbatively different that breaks down at very late times, and when it does it is \pref{INTpictureNZ} that is the more reliable equation.   
\end{itemize}

Since $\boldsymbol{\varrho}$ is a Hermitian $2\times 2$ matrix with unit trace, its elements $\varrho_{21} = \varrho_{12}^\ast$ and $\varrho_{22} = 1 - \varrho_{11}$ can be eliminated from \pref{INTpictureNZ} to leave the following two decoupled evolution equations for the remaining two variables $\varrho_{11}$ and $\varrho_{12}$:
\bea \label{NZ11}
\frac{\partial \varrho^{\ssI}_{11}}{\partial \tau} &=& g^{2} \int_{-\tau}^{\tau} \exd s \ e^{- i \omega s} G_\Omega(\tau,  \tau - s)  \\
&& \qquad \qquad \qquad  - 4 g^2 \int_0^{\tau} \exd s\ \mathrm{Re}\big[  G_\Omega(\tau,  \tau - s)  \big] \cos(\omega s) \varrho^{\ssI}_{11}(\tau- s)\,,\nn
\eea
and
\begin{eqnarray}\label{NZ12}
\frac{\partial \varrho^{\ssI}_{12}}{\partial \tau}  & = & - i  \delta\omega \;\varrho^{\ssI}_{12}(\tau) - 2 g^2 \int_0^{\tau} \exd s\ \mathrm{Re}\big[ G_\Omega(\tau,  \tau - s) \big] e^{+ i \omega s} \varrho^{\ssI}_{12}(\tau - s)  \\
&& \qquad\qquad\qquad\qquad + 2 g^2 e^{+ 2 i \omega \tau} \int_0^{\tau} \exd s\ \mathrm{Re}  \big[ G_\Omega(\tau,  \tau - s) \big] e^{- i \omega s} \varrho^{\ssI \ast}_{12}(\tau - s) \,. \nn
\end{eqnarray}
These two equations perform a change of integration variables $s \to \tau - s$ relative to \pref{INTpictureNZ}, since the result takes a particularly simple form when the Wightman functions are translation invariant in $\tau$. These are the main equations on which the remainder of the paper rely.

We note in passing that it can happen that the appearance in the above equations of the oscillatory factors $e^{\pm i\omega s}$ and $e^{i\omega\tau}$ can complicate the construction of their solutions. Such terms can be removed from an ordinary differential equation by standard changes of dependent variable, which in the present instance amount to returning to the Schr\"odinger picture. The result in Schr\"odinger picture is
\begin{eqnarray}\label{SchPicNZ11}
\frac{\partial \varrho_{11}}{\partial \tau} & = & g^{2} \int_{-\tau}^{\tau} \exd s \ e^{- i \omega s} G_\Omega(\tau,  \tau - s)  \\
& \ & \quad \quad \quad \quad \quad \quad \quad \quad - 4 g^2 \int_0^{\tau} \exd s\ \mathrm{Re}\big[  G_\Omega(\tau,  \tau - s)  \big] \cos(\omega s) \varrho^{\ssI}_{11}(\tau- s) \,, \nn
\end{eqnarray}
and
\begin{eqnarray}\label{SchPicNZ12}
\frac{\partial \varrho_{12}}{\partial \tau}  & = & - i ( \omega + \delta \omega ) \varrho_{12}(\tau) - 2 g^2 \int_0^{\tau} \exd s\ \mathrm{Re}\big[ G_\Omega(\tau,  \tau - s) \big] \varrho_{12}(\tau - s) \\
& \ & \qquad \qquad \qquad \qquad \quad \quad \quad \quad + 2 g^2 \int_0^{\tau} \exd s\ \mathrm{Re} \big[ G_\Omega(\tau,  \tau - s) \big] \varrho^{\ast}_{12}(\tau - s) \,.\notag
\end{eqnarray}

\subsection{Near-horizon Wightman function}

So far the description of qubit evolution has been quite general, with little said about the specific field state $\Omega$ or about the details of the qubit trajectory. Application of this formalism to a qubit near a black hole requires filling in some of this detail, starting with some information about the scalar-field Wightman function in a Schwarzschild geometry.

\subsubsection*{Hadamard correlation functions}

As mentioned earlier, for generic trajectories the scalar field Wightman function can be quite complicated, even for comparatively simple states like the Hartle-Hawking, Unruh or Boulware vacua \cite{Candelas:1980zt, Page:1982fm, Candelas:1984pg, Matyjasek:1998mq, Boulware:1974dm, DeWitt:1975ys, Hartle:1976tp, Unruh:1976db, Christensen:1977jc}. One of our central points is that the late-time evolution very close to the horizon does not depend on which of these choices for field state is made, with universal predictions relying only on the state being `Hadamard', in the sense that the Wightman correlation function 
\be
  G_\Omega (x,x') := \underset{\phi}{\mathrm{Tr}}\Bigl[ \Omega \; \phi(x) \, \phi(x') \Bigr] 
\ee
has -- in four spacetime dimensions -- the following limit as $x \to x'$ \cite{Fulling:1978ht,Kay:1988mu,Wald:1995yp,Radzikowski:1996pa,Radzikowski:1996ei}:
\be \label{hadamard}
 G_\Omega (x,x') = \frac{1}{8\pi^2} \left\{ \frac{ { \Delta^{1/2}(x,x') } }{ \sigma_{\epsilon}(x,x') } + V(x,x') \log\left| \frac{ \sigma_{\epsilon}(x,x') }{L^2} \right| + W_{\Omega}(x,x') \right\}\ , 
\ee
with
\begin{eqnarray} \label{cTdef}
\sigma_{\epsilon}(x,x') & := & \sigma(x,x') + 2 i \epsilon [ \cT(x) - \cT(x') ] + \epsilon^2 \ ,
\end{eqnarray}
and $\sigma(x,x')$ the so-called Synge world function \cite{J.L.Synge:1960zz,DeWitt:1960fc} that is equal to half the square of the geodetic length between $x$ and $x'$ (see Appendix \ref{App:Synge}). Here $\cT$ is any future-increasing function of time, and $\epsilon \to 0^+$ a small-distance regulator with dimensions of length that appears in the above formula so that $G_\Omega (x,x')$ satisfies the correct temporal boundary conditions. 

The quantities $\Delta(x,x')$, $V(x,x')$ and $W_{\Omega}(x,x')$ are biscalar functions that are symmetric in $x \leftrightarrow x'$, and regular in the limit that $x \to x'$. The renormalization length scale $L>0$ is introduced on dimensional grounds, and different values for $L$ can be absorbed into the precise definition of $W_{\Omega}(x,x')$. The subscript $\Omega$ on $W_\Omega$ is meant to emphasize that its detailed form depends on the state ${\Omega}$ \cite{Hack:2012qf}. The same is {\it not} true of the functions $\Delta(x,x')$ and $V(x,x')$, which are universal in the sense that they depend only on the geometry of the spacetime (and -- in the case of $V(x,x')$ -- on parameters like the mass of the field). 

What this says is that the leading part of the coincident limit of $G_\Omega(x,x')$ is universal in curved space, and shares in particular the singularity structure also found in flat space. The Hadamard form expresses the physical condition common to all effective field theories \cite{TheBook} that states that the details of very high-energy field modes are irrelevant provided because for slowly changing backgrounds they are prepared within their adiabatic vacuum. This amounts to a quantum variant of the principle of equivalence: modes with wavelengths much shorter than the local radius of curvature do not `know' that they are in curved space.

Because they depend only on local properties, there is a general procedure for computing the geometric functions $V(x,x')$ and $\Delta(x,x')$ in the coincident limit, for which $\sigma(x,x') \to 0$  \cite{DeWitt:1960fc, Decanini:2005gt}. For a real massive scalar field  evaluated on a Ricci-flat spacetime (like the Schwarzschild geometry) they have the form  
\begin{eqnarray}
\Delta^{1/2}(x,x')  & = & 1  +  \frac{1}{360} R^{\alpha\ \beta}_{\ \mu \ \nu} R_{\alpha \lambda \beta \rho}\, \sigma^{\mu} \sigma^{\nu} \sigma^{\lambda} \sigma^{\rho} +   \mathcal{O}( \sigma^{5/2} ) \\
V(x,x') & = &   \left( \frac{m^2}{2} - \frac{1}{360} R^{\rho\sigma\tau}_{ \ \ \ \mu} R_{\rho \sigma \tau \nu} \sigma^{\mu}\sigma^{\nu} \right) + \left( \frac{m^4}{16} + \frac{1}{1440} R_{\rho\sigma\tau\kappa} R^{\rho \sigma \tau \kappa} \right) \sigma_{\mu} \sigma^{\mu} + \mathcal{O}(\sigma^{3/2})\,,\nn
\end{eqnarray}
where $\partial_{\mu} \sigma = \sigma_{\mu}$ and $\sigma^\mu = g^{\mu\nu} \sigma_\nu$ obey the relation $\sigma_{\mu}\sigma^{\mu} = 2 \sigma$. Terms written $\mathcal{O}(\sigma^{3/2})$ are those containing three or more factors of $\sigma_{\mu}$.  For massive fields it is conventional to choose the form of $W_{\Omega}(x,x')$ so that $L^2 =  2 / m^2$, so that
\be \label{hadamardmass}
 G_\Omega (x,x') \simeq \frac{1}{8\pi^2} \left\{ \frac{  1 }{ \sigma_{\epsilon}(x,x') } + \left( \frac{m^2}{2} + \ldots \right) \log\left| \frac{m^2 \sigma_{\epsilon}(x,x') }{ 2} \right| +\cdots \right\}\,.
\ee

Of the vacuum states described above, the Hartle-Hawking \cite{Sanders:2013vza} and Unruh vacua \cite{Dappiaggi:2009fx} are both Hadamard states, and so share the same values for $\Delta(x,x')$ and $V(x,x')$ but not for $W_{\Omega}(x,x')$). The Boulware vacuum is not, however, as can be seen from its singular form for the stress-energy tensor at the horizon \cite{Christensen:1977jc}.

In practice the leading behaviour suffices for our purposes, which means we may use $\Delta(x,x') \simeq 1$ and drop $V(x,x')$ in the applications to follow, leaving the result
\begin{eqnarray}\label{HadSch}
 G_\Omega(x,x') \simeq \frac{ 1 }{8 \pi^2 \left[ \sigma(x,x') - i \epsilon [ \cT(x) - \cT(y) ] + \epsilon^2 \right]}  + \cdots\,,
\end{eqnarray}
that applies when $|\sigma(x,x')|$ is much smaller than both $r_s^2$ and $m^{-2}$. 

\subsubsection*{Evaluation for qubits hovering near the horizon}

What is special about the small-$\sigma(x,x')$ limit is that it applies not just as $x \to x'$, but also when $x$ and $x'$ are generic points situated sufficiently close to a null geodesic. Small $\sigma(x,x')$ should apply in particular for any two points hovering at a fixed position $(r,\theta,\phi) = (r_0, \theta_0, \phi_0)$ just outside the Schwarzschild event horizon, with $\sigma(x,x') \to 0$ as $r_0 \to r_s$. 

The function $\sigma(x,x')$ is evaluated in this limit in Appendix \ref{App:Synge} for points on such a hovering trajectory as a function of their separation $\Delta t$ in Schwarzschild time, with the result \cite{Emelyanov:2018woe,Singha:2018vaj,Chatterjee:2019kxg})
\be \label{wf5napp}
\sigma(x,x') = - 8 r_{s}^2 \left( 1 - \frac{r_{s}}{r_0} \right) \sinh^{2}\left( \frac{\Delta t}{4r_{s}} \right)  + \mathcal{O} \left( \frac{\sigma^{2}}{r_s^2} \right)
\ee
in the limit $\sigma(x,x') \to 0$. It is important that \pref{wf5napp} remains valid even if $\Delta t \gg r_s$, provided that $r_0$ is chosen close enough to $r_s$ to ensure that $|\sigma(x,x')| \ll r_s^2$. (The validity of this approximation in the regime $\Delta t \gg r_s$ is verified numerically in Appendix \ref{App:Synge}.)

For separations for which \pref{wf5napp} applies, eq.~\pref{HadSch} states that the Wightman function for any Hadamard state has the form 
\be  \label{WS}
G_\Omega(t+\Delta t, t)  \simeq  - \; \frac{1}{64\pi^2 r_{s}^2 \left( 1 - \frac{r_{s}}{r_0} \right) \Bigl(  \sinh\left[ {\Delta t }/({4 r_s}) \right] - i {\epsilon}/({4 r_{s}) } \Bigr)^2} + \cdots \,.
\ee

\section{Universal late-time near-horizon evolution}
\label{sec:Universal}

This section ties everything together to obtain a closed-form expression for the two universal thermalization time-scales that arise for qubits hovering asymptotically close to the horizon. The result is surprisingly simple because of an apparently paradoxical result: the simplicity occurs because in the near-horizon limit one can exploit the Wightman function's small-$\sigma(x,x')$ Hadamard form \pref{hadamard}. This seems paradoxical because thermalization occurs in the limit of very long time separations, $\Delta t \gg r_s$. The coexistence of these two limits is possible only because of the enormous time-dilation that relates static clocks running very near the horizon and those far from the black hole; two near-horizon events separated by a small geodesic separation can look to a distant observer  like they are separated by very large times.  

\subsection{The near-horizon Nakajima-Zwanzig equation}

The starting point is the interaction-picture Nakajima-Zwanzig equations \pref{NZ11} and \pref{NZ12} for the qubit state $\boldsymbol{\varrho}^{\ssI}(\tau)$. At order $g^2$ this gives 
\be 
\frac{\partial \varrho^{\ssI}_{11}}{\partial \tau} = g^{2} \int_{-\tau}^{\tau} \exd s \ e^{- i \omega s}\, \mfF \left(\gamma s\right) - 4 g^2 \int_0^{\tau} \exd s \, \mathrm{Re}\left[  \mfF\left(\gamma s \right) \right] \,\cos(\omega s)\,  \varrho^{\ssI}_{11}(\tau - s) \,,
\ee
and
\begin{eqnarray}
\frac{\partial \varrho^{\ssI}_{12}}{\partial \tau}  & = & - i  \delta\omega \;\varrho^{\ssI}_{12}(\tau) - 2 g^2 \int_0^{\tau} \exd s\ \mathrm{Re}\left[  \mfF \left(\gamma s \right) \right] e^{+ i \omega s} \varrho^{\ssI}_{12}(\tau - s)  \\
&& \qquad\qquad\qquad\qquad + 2 g^2 e^{+ 2 i \omega \tau} \int_0^{\tau} \exd s\ \mathrm{Re}\left[  \mfF\left(\gamma s \right) \right] e^{- i \omega s} \varrho^{\ssI \ast}_{12}(\tau - s) \,, \nn
\end{eqnarray}
where 
\be
  \mfF( \Delta t) := G_\Omega(t + \Delta t, t) = \mfF(\gamma \Delta \tau) \quad \hbox{with} \quad \gamma := \frac{1}{\sqrt{1-r_s/r_0}} \,.
\ee

Our later interest is in late times as seen by an observer far from the black hole, so changing coordinates $\tau = t/\gamma$ gives
\be \label{NZ11t}
\frac{\partial \varrho^{\ssI}_{11}}{\partial  t}  = g^{2} \int_{- t}^{t} \exd s \ e^{- i \wred s} \ol{\mfF}(s) - 4 g^2 \int_0^{ t} \exd s\ \mathrm{Re}\left[  \; \ol{\mfF}(s) \right] \cos(\wred s) \varrho^{\ssI}_{11}(t - s) \,,
\ee
and
\begin{eqnarray} \label{NZ12t}
\frac{\partial \varrho^{\ssI}_{12}}{\partial t}  & = & - i  \delta \wred \;\varrho^{\ssI}_{12}(t) - 2 g^2 \int_0^{t} \exd s\ \mathrm{Re}\left[  \; \ol{\mfF}(s) \right] e^{+ i \wred s} \varrho^{\ssI}_{12}(t - s) \\
&& \qquad\qquad\qquad\qquad + 2 g^2 e^{+ 2 i \wred  t} \int_0^{t} \exd s\ \mathrm{Re}\left[  \; \ol{\mfF}(s) \right] e^{- i \wred s} \varrho^{\ssI \ast}_{12}( t - s) \,,\nn
\end{eqnarray}
where for convenience we define the redshifted qubit gap as seen by observers looking at the qubit far from the black hole
\be
\wred  :=  \omega \sqrt{ 1 - \sfrac{r_{s}}{r_0} }   \ ,
\ee
and perform a similar scaling of the $\mathcal{O}(g^2)$ counter-term $\delta\omega_{\infty} := \delta \omega ( 1 - r_{s} / r_{0} )$. Finally, $\ol{\mfF}$ denotes the scaled Wightman function 
\begin{eqnarray}
\ol{\mfF}( t )  :=  \left( 1 - \frac{r_{s}}{r_0} \right) \mfF( t)  \,.
\end{eqnarray}

In the small-$\sigma(x,x')$ limit inspection of \pref{WS} shows that $\ol{\mfF}$ has the simple asymptotic form
\be \label{NearHorBarF}
\ol{\mfF}( \Delta t )  \simeq  -\;  \frac{1}{64\pi^2 r_{s}^2 \left( \sinh\left[ {\Delta t }/{(4 r_s)} \right] - i \epsilon / (4r_{s}) \right)^2 } \ ,
\ee
which is identical to the analogous result for the massless Rindler correlation function found in \cite{Kaplanek:2019dqu} once one replaces $r_{s} \to 1 / (2 a)$. Recall from Appendix \ref{App:Synge} that this asymptotic form for $\ol{\mfF}(t)$ is valid so long as $|\sigma(x,x') | \ll r_s^2$ and so applies when
\begin{eqnarray}
 \frac{\Delta t}{r_{s}} \ \ll \ \left| 2 \log\left( \frac{1 - r_{s} / r_0}{4} \right) \right| \,,
\end{eqnarray}
and so in particular there is always an $r_0 > r_s$ sufficiently close to the horizon for which this is satisfied, no matter how large $\Delta t/r_s$ happens to be.

\subsection{The late-time Markovian approximation}

From here on the story evolves much as it did in the Rindler example considered in \cite{Kaplanek:2019dqu}, by virtue of the similarity between \pref{NearHorBarF} and its counterpart for an accelerated qubit in flat spacetime. 

In particular eqs.~\pref{NZ11t} and \pref{NZ12t} greatly simplify when $\boldsymbol{\varrho}^{\ssI}$ is slowly varying (compared with the light-crossing time of the black hole, $r_s$) and we focus on $t \gg r_s$, because in this case the sharply peaked form for $\ol{\mfF}(t)$ allows the upper integration limit to be taken to infinity, and implies that a Taylor expansion in the integrand of $\varrho_{ij}(t-s)$ in powers of $s$ converges very quickly. 

After choosing $\delta \omega$ to cancel the field-induced shift in qubit energy -- which means picking 
\be \label{renormCT}
  \delta \omega_{\infty} = \delta \omega \left(1 - \frac{r_{s}}{r_0} \right) = - g^2 \shift \,,
\ee
with $\shift$ defined by (\ref{shift}) below --  these steps lead (at face value) to the following approximate evolution equations (see \cite{Kaplanek:2019dqu} for details) 
\be \label{M11t}
\frac{\partial \varrho^{\ssI}_{11}( t)}{\partial  t}  \simeq \frac{2g^2 \CS}{e^{4 \pi r_s \wred} + 1} - 2 g^2 \CS \, \varrho^{\ssI}_{11}( t )    \,,
\ee
and
\be \label{M12t}
\frac{\partial \varrho^{\ssI}_{12}( t)}{\partial t}    \simeq    - g^2 \CS \,\varrho_{12}^{\ssI}( t) + g^2 ( \CS - i \shift )\, e^{+2 i \wred t}\varrho_{12}^{\ssI\ast}(t) \,,
\ee
in which the quantities $\CS$ and $\shift$ are defined by
\be\label{Cdef}
\CS =  2 \int^{\infty}_0 \exd s \, \mathrm{Re}[\ol{\mfF}(s)] \cos(\wred s) \ \simeq \  \frac{\wred \coth\left( 2 \pi r_s \wred \right) }{4\pi} 
\ee
and\footnote{A flat-space analog of $\shift$ is computed in \cite{Kaplanek:2019dqu} for generic field masses $m \neq 0$ and with the replacement $a \to {2}/{r_{s}}$ (using the different notation $\Delta_{\ssM}$ there). Equation (\ref{shift}) follows as the $m \to 0^{+}$ limit of this function (this same function is evaluated in \cite{Moustos:2016lol}).}
\be \label{shift}
\shift =  2  \int^{\infty}_0 \exd s \ \mathrm{Re}[\ol{\mfF}(s)] \sin(\wred s) \ \simeq \   \frac{\wred}{2 \pi^2} \log\left( \frac{e^{\gamma} \epsilon}{2r_s} \right) + \frac{\wred}{2 \pi^2} \mathrm{Re}\left[ \psi^{(0)}( - 2 i r_s \wred ) \right] \,.
\ee
where $\psi^{(0)}(z) = \Gamma'(z) / \Gamma(z)$ is the digamma function \cite{NIST}.

\subsubsection*{Control over approximations}

The words `at face value' are added above eqs.~\pref{M11t} and \pref{M12t} because the term involving $\cD_\ssS$ must actually be dropped in the above if we are consistent. The reasons for this lie in the size of the deviations from the leading approximation, and the assumptions that must be made in order to neglect them. We briefly summarize the issues, following closely the discussion in \cite{Kaplanek:2019dqu, Kaplanek:2019vzj}. A side effect of this observation -- together with the energy shift \pref{renormCT} -- is the elimination of all singular dependence\footnote{For the purposes of estimating the size of different contributions we take $\epsilon$ here to be much smaller than other scales, but not infinitely small so that logarithms of $\epsilon$ cannot overwhelm powers of $g^2$.} in the limit $\epsilon \to 0$ that enters through eq.~\pref{shift}. 

There are two kinds of approximations to consider -- one convenient and one essential. The issue of convenience concerns the relative size of the qubit splitting $\omega$ and the generic size of field-driven corrections to this splitting. Assuming $\omega_\infty$ to be much larger than the corrections to $\omega$ induced by the interactions with the field simplifies calculations by allowing use of non-degenerate methods. In terms of the functions $\cC_\ssS$ and $\cD_\ssS$ this condition requires 
\be
    \frac{g^2 \CS}{\wred}  \ll 1 \quad \hbox{and} \quad  \;  \frac{g^2 \shift}{\wred}   \ll 1 \,.
\ee
Table \ref{MarkovianTable0} displays the asymptotic form for these two quantities in the limit of large and small $\omega_\infty r_s$, showing that they require $\omega_\infty r_s$ not to be taken smaller than $g^2 / 4\pi$. 

\begin{table}[h] 
\centering    
\centerline{\begin{tabular}{ r|c|c| }
\multicolumn{1}{r}{}
& \multicolumn{1}{c}{$\underset{\ }{ {g^2 \CS}/{\wred} \ll 1 }$}
& \multicolumn{1}{c}{${g^2 \shift}/{\wred} \ll 1$} \\
\cline{2-3}
$r_s \wred \ll 1$ & $\stackrel{\ }{ \underset{\ }{ \frac{ g^2 }{ 8 \pi^2 r_s \wred }  } }$ & $\stackrel{\ }{\frac{g^2}{2\pi^2}\log\left[ {\epsilon}/{(2r_{s})} \right]}$     \\
\cline{2-3}
$r_s \wred \gg 1$ & $\stackrel{\ }{ \underset{\ }{ \frac{ g^2 }{ 4 \pi  }  } }$ & $\frac{g^2}{2\pi^2} \log(e^{\gamma} \wred \epsilon)$    \\ 
\cline{2-3} 
\end{tabular} }
      \caption{The large- and small-$\omega_\infty r_s$ asymptotic forms for the two quantities that must be small to work with nondegenerate perturbation theory (see \cite{Kaplanek:2019dqu}).} \label{MarkovianTable0}
\end{table}

The essential approximation is the one that makes the Markovian evolution dominate the Nakajima-Zwanzig evolution. To see what this involves, recall that the Markovian approximation is derived from the Nakajima-Zwanzig equation by Taylor expanding $\varrho_{ij}^{\ssI}(t - s) \simeq \varrho^\ssI_{ij}(t) - s \dot{\varrho}^\ssI_{ij}(t) + \cdots$ inside the integrands of equations (\ref{NZ11t}) and (\ref{NZ12t}),
\begin{eqnarray} \label{TaylorEq}
g^2 \int_{0}^{t} \exd s\ f(s) \varrho^\ssI_{ij}(t - s) &  \simeq & g^2 \int_{0}^{\infty} \exd s\ f(s) \big[ \varrho^\ssI_{ij}(t) - s \dot{\varrho}^\ssI_{ij}(t) + \ldots \big]
\end{eqnarray}
where $t \gg r_{s}$ is used to take the upper limit of integration to infinity (given the exponential falloff of $f(s)$ for $s \gg r_s$). The size of the $s\dot{\varrho}^\ssI_{ij}(t)$ term characterizes the size of deviations from the Markovian limit, and we evaluate it to understand what demands are made on the free parameters of the model by the requirement that these be small. Physically this amounts to requiring the evolution time-scale of $\varrho^\ssI_{ij}$ to be large compared with the domain of support of the rest of the integrand. 

The quantitative conditions are obtained self-consistently, by evaluating $\dot\varrho^\ssI_{ij}$ assuming the time dependence is given by \pref{M11t} and \pref{M12t}, whose integration implies
\be 
\varrho_{11}^{\ssI}(t) = \frac{1}{e^{4 \pi r_s \wred} + 1} + \left[ \varrho_{11}(0) - \frac{1}{e^{ 4 \pi r_s \wred } + 1 } \right] e^{ - 2 g^2 \cC_\ssS\,  t } \,, \label{M11solx}
\ee
and
\be 
\varrho_{12}^{\ssI}(  t) = e^{ -  g^2 \cC_\ssS \, t  } \left[ \varrho_{12}(0) + \varrho_{12}^{\ast}(0) \left( \frac{g^2 \shift}{2\wred} + i \frac{g^2 \CS}{2\wred} \right) (1 - e^{ 2 i \wred   t }) \right] \,. \label{M12solx}
\ee
Differentiating this to find $\dot\varrho^\ssI_{ij}$ then allows the $\dot\varrho^\ssI$ term to be computed in equations like \pref{TaylorEq}, and requiring the result to be negligible relative to the leading term for all of the integrals appearing in eqs.~\pref{NZ11t} and \pref{NZ12t} requires the following four quantities all to be negligible: 
\be
 g^2 \frac{\mathrm{d} \CS}{\mathrm{d} \wred} \ll 1 \,, \quad 
  g^2 \frac{\mathrm{d} \shift}{\mathrm{d} \wred} \ll 1 \,, \quad
   \frac{\wred}{\CS} \frac{\mathrm{d} \CS}{\mathrm{d} \wred} \ll 1  \quad \hbox{and} \quad
 \frac{\wred}{\CS} \frac{\mathrm{d} \shift}{\mathrm{d} \wred}   \ll 1 \,.
\ee
The first two of these are required whenever the derivative in $\dot\varrho_{ij}$ is of order $g^2 \cC_\ssS$ while the second two arise when it is order $\omega_\infty$. [Differentiation with respect to $\omega_\infty$ arises from use of identities like $s \cos( \omega_\infty s) = (\exd/\exd\omega_\infty) \sin(\omega_\infty s)$ in equations like \pref{TaylorEq}.]

\begin{table}[h] 
\centering    
\centerline{\begin{tabular}{ r|c|c|c|c| }
\multicolumn{1}{r}{}
& \multicolumn{1}{c}{$\underset{\ }{ g^2 \frac{\mathrm{d} \CS}{\mathrm{d} \wred} \ll 1 }$} 
& \multicolumn{1}{c}{$g^2 \frac{\mathrm{d} \shift}{\mathrm{d} \wred} \ll 1$}
& \multicolumn{1}{c}{$\frac{\wred}{\CS} \frac{\mathrm{d} \CS}{\mathrm{d} \wred} \ll 1$}
& \multicolumn{1}{c}{$\frac{\wred}{\CS} \frac{\mathrm{d} \shift}{\mathrm{d} \wred} \ll 1$} \\
\cline{2-5}
$r_s \wred \ll 1$   & ${g^2 r_s \wred}/{3}$  & $\frac{g^2}{2\pi^2}\log\left[ {\epsilon}/{(2r_{s})} \right]$ & ${8\pi^2 r_s^2 \omega_{\infty}^2}/{3}$  & $4 r_{s} \wred \log\left[ {\epsilon}/{(2r_{s})} \right]$  \\
\cline{2-5}
$r_s \wred \gg 1$   & ${ g^2 }/{ 4 \pi }$  & $\frac{g^2}{2\pi^2}\log(e^{\gamma} \wred \epsilon)$   & $1$  & $\frac{2}{\pi} \log( e^{\gamma}\wred \epsilon )$  \\ 
\cline{2-5} 
\end{tabular} }
      \caption{The large- and small-$\omega_\infty r_s$ asymptotic forms for the four quantities that must be small to believe the Markovian approximation to the Nakajima-Zwanzig equation (see \cite{Kaplanek:2019dqu}). Notice that $r_{s} \wred \gg 1$ is incompatible with Markovian evolution.} \label{MarkovianTable}
\end{table}

Table \ref{MarkovianTable} displays the asymptotic behaviour for these four quantities in the limits where $r_{s}\wred$ is very large or very small. This table makes clear in particular that only $r_{s} \wred \ll 1$ is consistent in the Markovian regime, since otherwise the bounds ${\wred \CSp}/{\CS}  \ll 1$ and ${\wred \cD_\ssS^{\prime}}/{\CS}  \ll 1$ necessarily fail (because the $e^{i\omega_\infty t}$ oscillations are too rapid). 

The asymptotic forms of Table \ref{MarkovianTable} say more than just this, however. From them we also notice that $r_{s} \wred \ll 1$ implies\footnote{Using $\psi^{(0)}(z) \simeq \frac{1}{z} - \gamma + \frac{\pi^2 z}{6} - \zeta(3) z^2 + \ldots$ for $|z|\ll 1$ (with $\zeta$ the Riemann zeta function) \cite{NIST}, $\shift \simeq \frac{\wred}{2\pi^2} ( \log( \frac{\epsilon}{2 r_{s}} ) + 4 \zeta(3) (r_s\wred)^2  + \mathcal{O}[(r_s\wred)^4] )$ and $\shiftp \simeq \frac{1}{2\pi^2} ( \log( \frac{\epsilon}{2 r_{s}} ) + 12 \zeta(3) (r_s\wred)^2 + \mathcal{O}[ (r_s\wred)^4 ] ).$ } $\shift / \omega \sim \shiftp$ and so
\begin{eqnarray}
\frac{g^2 \shift}{\omega_\infty} \simeq g^2 \frac{\exd \shift}{\exd \wred}  = \left( \frac{g^2 \CS}{\wred} \right) \times \left( \frac{\wred}{\CS} \frac{\exd \shift}{\exd \wred} \right) \ll \frac{g^2 \CS}{\wred} \ ,
\end{eqnarray}
which implies that the ${g^2 \shift}/{\omega_\infty}$ term appearing in the solution \pref{M12solx} is negligible relative to the ${g^2 \CS}/{\omega_\infty}$ term. This means that the $g^2\shift$ terms in the Markovian evolution equations \pref{M12t} can be neglected, allowing the Markovian evolution instead to be written as \pref{M11t} and
\be \label{M12t'}
\frac{\partial \varrho^{\ssI}_{12}( t)}{\partial t}    \simeq    - g^2 \CS \,\varrho_{12}^{\ssI}( t) + g^2  \CS  \, e^{+2 i \wred t}\varrho_{12}^{\ssI\ast}(t) \,.
\ee
In particular the divergent quantity $\shift$ plays no role in the Markovian limit, apart from shifting the qubit energy levels in the way that is renormalized into the definition of $\omega$. Following the steps discussed at great length in \cite{Kaplanek:2019dqu,Kaplanek:2019vzj}) shows that these equations preserve positivity of $\boldsymbol{\varrho}(t)$ to $\mathcal{O}(g^2)$ in the Markovian limit, with no additional approximations necessary. 

The solutions in the Markovian regime therefore become
\be 
\varrho_{11}^{\ssI}(t) = \frac{1}{e^{4 \pi r_s \wred} + 1} + \left[ \varrho_{11}(0) - \frac{1}{e^{ 4 \pi r_s \wred } + 1 } \right] e^{ - 2   t / \xi } \,, \label{M11sol}
\ee
and
\be 
\varrho_{12}^{\ssI}(  t) = e^{ -   t / \xi } \left[ \varrho_{12}(0) +  i \varrho_{12}^{\ast}(0) \frac{g^2 \CS}{2\wred} (1 - e^{ 2 i \wred   t }) \right] \,, \label{M12sol}
\ee
where
\begin{eqnarray}
\xi \ := \ \frac{1}{g^2 \CS} \ = \ \frac{4\pi \tanh\left( 2 \pi r_s \wred \right) }{g^2 \wred} \ \simeq \ \frac{8\pi^2 r_s}{ g^2  } 
\end{eqnarray}
and the last line follows since the Markovian approximation demands $\wred r_{s} \ll 1$. These solutions describe the exponential decay towards a thermal distribution (with temperature $T = 1/(4\pi r_s) = T_\ssH$ that equals the Hawking temperature), doing so with the characteristic time-scale $\xi \simeq 8\pi^2 r_s/g^2$. Notice that the approach to equilibrium takes place twice as fast for the diagonal components of $\boldsymbol{\varrho}$ compared to its off-diagonal parts. 

We remark in passing that it is also possible to solve the Nakajima-Zwanzig equation at late times using weaker assumptions than those that lead to the above Markovian solutions, using methods similar to those used in \cite{Kaplanek:2019vzj} (see also \cite{Moustos:2016lol}, where a non-Markovian solution for an accelerated qubit is derived by method of Laplace transforms). The utility of such a solution is less interesting here since the Markovian condition $r_{s}\wred = r_{s} \omega \sqrt{ 1 - r_{s} / r_0 } \ll 1$ is satisfied for any qubit of fixed rest-frame energy splitting that hovers sufficiently close to the black-hole horizon.

\subsubsection*{Frame Independence of the Markovian Limit}

Since the solution (\ref{M11sol}) and (\ref{M12sol}) refers to the redshifted time $\Delta t$ defined in (\ref{timerelation}) one may wonder about the physical meaning of tracking a time coordinate as measured by a hovering observer far from the black hole horizon, and whether the discussion of perturbativity and Markovianity applies only in such a frame.

What matters is that there is a hierarchy of scales between the late times of interest, $\Delta t(r)$, (such as the equilibration time) and the correlation time of the environment, $\tau_c(r)$ (in this case, the local Hawking temperature). These can equally well be compared at the position of the qubit or by an observer at infinity. Although gravitational redshift changes both $\Delta t$ and $\tau_c$ (and this is why they depend on $r$), this redshift cancels in their ratio; observers at all radii agree that $\Delta t(r) \gg \tau_c(r)$. Both a local observer travelling with the qubit and one hovering at spatial infinity agree on the heirarchies of scale
\bea
\frac{\Delta t}{r_{s}} \sim T_{\ssH} \Delta t \gg 1 \quad \quad \Longleftrightarrow \quad \quad \frac{\Delta \tau}{r_{s} \sqrt{ 1 - r_s/r_0}} \sim T(r_{0}) \Delta \tau \gg 1  \label{hierarchy}
\eea
where $T_{\ssH} = ( 4 \pi r_{s} )^{-1}$ is the Hawking temperature, and $T(r_{0}) = T_{\ssH}  / \sqrt{1 - r_{s} / r_0}$ is the local temperature. This means that there is nothing special about tracking the qubit evolution in terms of the redshifted time $\Delta t$, and in fact an observer at any radius will agree on a hierachy of scales as in (\ref{hierarchy}).

\section{Conclusions}

In summary, this paper shows how Open EFT methods can lend themselves to late-time resummation in more general gravitational systems than the cosmological examples previously explored. As in the examples of \cite{Kaplanek:2019dqu, Kaplanek:2019vzj} simplicity arises near the horizon at late times, even when the underlying geometry tends to makes quantum mechanical calculations difficult. Standard tools for open quantum systems give relatively easy access to times of the order $r_{s} / g^2$, at least in the specific instance of an Unruh-DeWitt detector placed very close to a Schwarzschild horizon and interacting with a quantum field. The resulting evolution describes qubit thermalization with the expected Hawking radiation, asymptoting to the Hawking temperature $T_{\ssH} = (4 \pi r_s)^{-1}$. The time-scale for thermalizing a hovering qubit can be computed, and in the very-near-horizon limit takes a universal form that relies only on properties of the near-horizon geometry given only the relatively weak assumption that the quantum field is prepared in a vacuum state of Hadamard form (including in particular the Hartle-Hawking and Unruh states).

What makes the late-time evolution easy to resum is its Markovian nature over Schwarzschild times that are long compared with $r_s$. Autocorrelations of the field in a Hadamard state then fall off very robustly for qubits hovering very near the horizon, effectively washing out the past entanglement history.  As one might expect from the equivalence principle, the qubit behaviour becomes equivalent to that of a qubit accelerating through flat space in the limit of infinite acceleration. It is the large acceleration (and blueshift) experienced by the qubit which ensures that the quantum field mass eventually becomes negligible in the near-horizon limit, explaining why the mass largely drops out of our result.  As a consequence, the Markovian evolution seems likely to be very robust, at least asymptotically close to the horizon (provided $r_{s} \wred \ll 1$, so that the qubit states are not too split to allow thermal excitation).

The absence of mass dependence (in the $m\, r_s \ll 1$ limit) also carries information about dependence on the scalar self-coupling, $\lambda$. Scalar self-couplings are known to give rise to secular effects for accelerated observers even in flat space \cite{Burgess:2018sou} (see also \cite{Akhmedov:2015xwa} for other evidence for secular growth in black-hole geometries), where it is known that they can also be resummed at late times. For the Rindler problem late-time resummation amounts to re-organizing perturbation theory using a small shifted mass, $\delta m^2 \sim \lambda a^2$, similar to the development of small temperature-dependent masses in thermal environments \cite{Burgess:2018sou}. Similarity with the Rindler problem makes it is very plausible that a similar resummation can be obtained near the Schwarzschild horizon by shifting the scalar mass by an amount $\delta m^2 \sim \lambda/r_s^2$, making the $m$-independence of near-horizon qubit evolution likely also to imply the same for $\lambda$-dependence, at least when $\lambda$ is small and times are late.

\section*{Acknowledgements}
We thank Viacheslav Emelyanov and Laszlo Zalavari for helpful discussions. This work was partially supported by funds from the Natural Sciences and Engineering Research Council (NSERC) of Canada. Research at Perimeter Institute is supported in part by the Government of Canada through the Department of Innovation, Science and Economic Development Canada and by the Province of Ontario through the Ministry of Colleges and Universities. 

\appendix

\section{The Synge world function}
\label{App:Synge}

This appendix derives some of the features of the Synge world-function that are used in the main text.  

\subsection*{Definitions} 

To this end consider two points, $x$ and $x'$, that are connected by a timelike geodesic $\Gamma$. If $\lambda$ is an affine parameterization of $\Gamma$ then it is described by the curve $y^\mu(\lambda)$ along which
\begin{eqnarray} \label{app:geoeq}
   \ddot{y}^{\mu}  + \Gamma^{\mu}_{\ \nu\sigma} \dot{y}^{\nu}  \dot{y}^{\sigma}  = 0
\end{eqnarray}
is obeyed for all $\lambda$, with $\dot y^\mu := \exd y^\mu/\exd \lambda$. The fact that $\Gamma$ connects $x$ and $x'$ is expressed as the boundary conditions $y(\lambda_{\mathrm{i}}) = x'$ and $y(\lambda_{\mathrm{f}}) = x$. 

For such a geodesic the Synge world function, $\sigma(x,x')$, is defined by \cite{J.L.Synge:1960zz,Poisson:2011nh,Ottewill:2008uu}
\begin{eqnarray}
\sigma(x,x') & := & \frac{1}{2} (\lambda_{\mathrm{f}} - \lambda_{\mathrm{i}} ) \int_{\lambda_{\mathrm{i}}}^{\lambda_{\mathrm{f}}} \exd \lambda \; g_{\mu\nu} \dot{y}^{\mu}  \dot{y}^{\nu}  \ , \label{app:worldfunction} 
\end{eqnarray}
where the integral is performed along the geodesic $\Gamma$. This integral is actually quite easy to evaluate because the geodesic equation \pref{app:geoeq} implies that the quantity 
\be
  \zeta :=  g_{\mu\nu}  \dot{y}^{\mu}  \dot{y}^{\nu}
\ee
is independent of $\lambda$ along $\Gamma$, and so $\sigma(x,x') = \frac12 \, \zeta \, (\lambda_{\mathrm{f}} - \lambda_{\mathrm{i}})^2$. For timelike curves $\zeta$ is negative, and if in that case the parameter is chosen to be proper time along the geodesic -- {\it i.e.}~if $\lambda = \tau$ -- then $\zeta = -1$ and $\lambda_{\mathrm{f}} - \lambda_{\mathrm{i}} = \Delta s$, establishing that
\be
    \sigma(x,x') = - \frac{1}{2} ( \Delta s )^2 , 
\ee
as used in the main text.

\subsection*{Expansion as $x \to x'$}

The dependence of $\sigma(x,x')$ on the geometry can be made explicit in the limit $x \to x'$. This is most easily done using \pref{app:worldfunction} and specializing the evaluation of $\zeta$ to the point $x'$ (as can be freely done since $\zeta$ is independent of $\lambda$), leading to
\begin{eqnarray}
\sigma(x,x') & = & \frac{1}{2} (\lambda_{\mathrm{f}} - \lambda_{\mathrm{i}})^2 \; g'_{\mu\nu}\dot{y}^{\mu}(\lambda_{\mathrm{i}}) \dot{y}^{\nu}(\lambda_{\mathrm{i}}) \,, \label{app:worldfunction3}
\end{eqnarray}
where here (and below) a prime on a field like $g'_{\mu\nu}$ indicates that it is evaluated at $x'$. 

Expanding $y^\mu(\lambda_{\rm f})$ in powers of $\lambda_{\rm f} - \lambda_{\rm i}$ gives
\be 
  y^{\mu}(\lambda_{\mathrm{f}})  =  y^{\mu}(\lambda_{\mathrm{i}}) +  (\lambda_{\mathrm{f}} - \lambda_{\mathrm{i}} ) \; \dot{y}^{\mu}(\lambda_{\mathrm{i}})  +  \frac{1}{2} (\lambda_{\mathrm{f}} - \lambda_{\mathrm{i}} )^2 \; \ddot{y}^{\mu}(\lambda_{\mathrm{i}})  +  \frac{1}{6} (\lambda_{\mathrm{f}} - \lambda_{\mathrm{i}} )^3 \; \dddot{y}^{\mu}(\lambda_{\mathrm{i}})  +  \ldots  , 
\ee
in which we use the boundary conditions $y(\lambda_{\mathrm{i}}) = x'$ and $y(\lambda_{\mathrm{f}}) = x$, as well as eliminating $\ddot y^\mu$ using the geodesic equation, leading to
\begin{eqnarray}
x^{\mu} - x^{\prime \mu} & = & (\lambda_{\mathrm{f}} - \lambda_{\mathrm{i}}) \; \dot{y}^{\mu}(\lambda_{\mathrm{i}})  -  \frac{1}{2} (\lambda_{\mathrm{f}} - \lambda_{\mathrm{i}})^2 \Gamma^{\mu\prime}_{ \rho\nu} \; \dot{y}^{\rho}(\lambda_{\mathrm{i}}) \dot{y}^{\nu}(\lambda_{\mathrm{i}}) \\
&  & \quad \quad \quad  -  \frac{1}{6}(\lambda_{\mathrm{f}} - \lambda_{\mathrm{i}})^3 \Bigl( \partial_{\rho} \Gamma^{\mu\prime}_{ \nu\sigma} - 2 \Gamma^{\mu\prime}_{ \rho\eta} \Gamma^{\eta\prime}_{ \nu \sigma} \Bigr) \; \dot{y}^{\rho}(\lambda_{\mathrm{i}}) \dot{y}^{\nu}(\lambda_{\mathrm{i}}) \dot{y}^{\sigma}(\lambda_{\mathrm{i}})  +  \ldots \,. \notag
\end{eqnarray}

Inverting the above gives a series expansion for $(\lambda_{\mathrm{f}} - \lambda_{\mathrm{i}}) \dot{y}^{\mu}(\lambda_{\mathrm{i}})$ in powers of $x - x'$:
\begin{eqnarray}
(\lambda_{\mathrm{f}} - \lambda_{\mathrm{i}} ) \dot{y}^{\mu}(\lambda_{\mathrm{i}}) & = & (x - x^\prime)^{\mu}   +   \frac{1}{2} \,\Gamma^{\mu\prime}_{ \lambda\nu}  (x - x')^{\lambda} (x - x')^{\nu} \\
&  & \quad \quad \quad  +  \frac{1}{6} \Bigl( \partial_{\lambda} \Gamma^{\mu\prime}_{ \nu\sigma} +  \Gamma^{\mu\prime}_{ \lambda\eta}\Gamma^{\eta\prime}_{ \nu\sigma}  \Bigr)  (x - x')^{\lambda} (x - x')^{\nu} (x - x')^{\sigma}  +  \ldots  \,,\notag  
\end{eqnarray}
which, when used in (\ref{app:worldfunction3}), gives \cite{J.L.Synge:1960zz,Ottewill:2008uu}
\be 
\sigma(x,x')   =  \frac{1}{2} g'_{\mu\nu}\; (x - x' )^{\mu} ( x - x')^{\nu} \ + \ \frac{1}{4} g'_{\mu\nu, \sigma} \; (x - x' )^{\mu} ( x- x')^{\nu}  ( x - x')^{\sigma} + \cdots \,. \label{wf4} 
\ee
One can continue in this way to any fixed order.\footnote{Although neither the coefficients nor $x-x'$ in this expansion are covariant, the final result is (transforming as a bi-scalar). A more explicitly covariant expression can be found by expanding in a more covariant variable, but expression \pref{wf4} suffices for our present purposes.}

\subsubsection*{Expansion for fixed $r$ in Schwarzschild}

We next evaluate \pref{wf4} for the special case $x$ and $x'$ lie along a trajectory at fixed $r = r_0$ (and $\theta$ and $\phi$) in the Schwarzschild geometry. Choosing $x'$ to correspond to $t_{\rm i} = 0$ and $x$ to be $t_{\rm f} = \Delta t$, we have (in Kruskal coordinates)
\bea 
  T - T'  &=&  \sqrt{\, \frac{r_0}{r_s} - 1 } \; \exp\left(\frac{r_0}{2r_s} \right) \sinh\left( \frac{\Delta t}{2r_{s}} \right)  \nn \\
  X - X' &=&   \sqrt{\, \frac{r_0}{r_s} - 1 } \;  \exp\left(\frac{r_0}{2r_s} \right) \left[ \cosh\left( \frac{\Delta t}{2r_{s}} \right) - 1 \right] \,.
\eea
So using (\ref{metricK}) 
\be
 -  g'_{\ssT\ssT} =  g'_{\ssX\ssX} =  \frac{4 r_{s}^3}{r_0} \; e^{ - r_{0} / r_{s}} \,,
\ee
the leading-order term in (\ref{wf4}) is \cite{Emelyanov:2018woe}
\bea \label{wf5}
\sigma(x,x')  & = & \frac{1}{2}\, g'_{\ssX\ssX} \; \Bigl[ - ( T  -  T' )^2 + (X - X')^2 \Bigr]   +  \cO[(x-x')^3] \nn\\
  &=& - 8 r_{s}^2 \left( 1 - \frac{r_{s}}{r_0} \right) \sinh^{2}\left( \frac{\Delta t}{4r_{s}} \right)  +  \cO[(x-x')^3] \ .  
\eea
which uses the identity $\sinh^2 a - ( \cosh a - 1 )^2 = 4 \sinh^2 ({a}/2)$.  

\begin{figure}[t]
\centering
\includegraphics[width=0.9\textwidth]{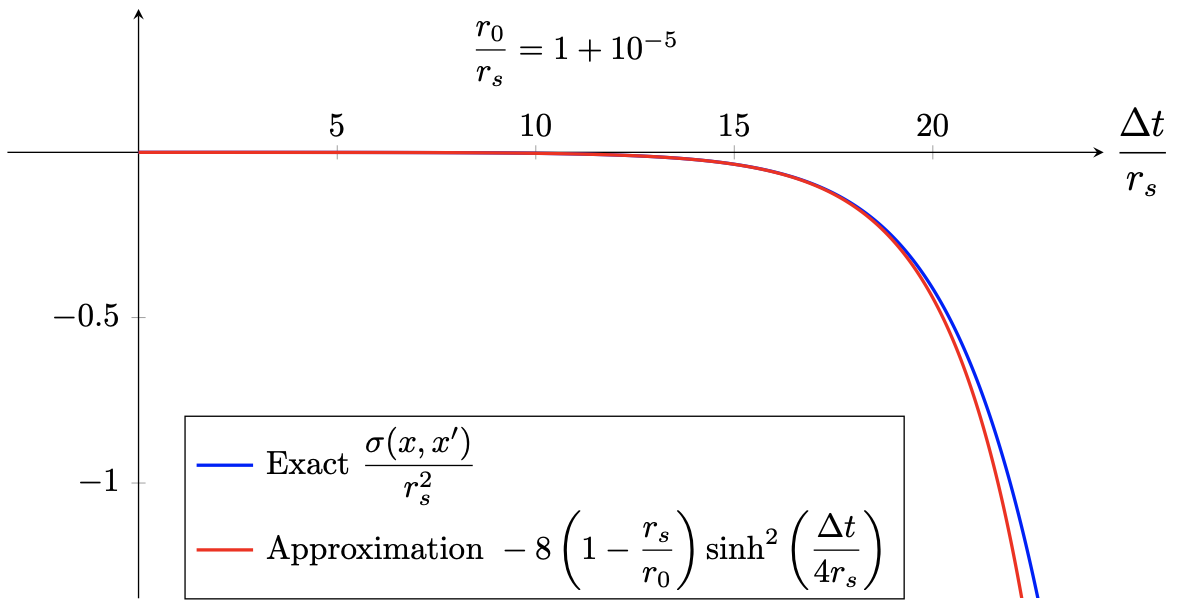}
\caption{\label{fig:Num1} Numerical comparison of the Synge world-function and the asymptotic expressions \pref{wf5} and \pref{wfSchwz}, showing how \pref{wf5} enjoys the broader domain of validity. This plot assumes $r_0/r_s = 1 + 10^{-5}$.}
\end{figure}

\vspace{2mm}
Evaluating the sub-leading terms in the series shows that corrections are of order
\be 
 \cO\left[(x-x')^3\right]] =  \mathcal{O}\left\{r_{s}^2 \left[ \left( 1 - \frac{r_{s}}{r_0} \right) \sinh^{2}\left( \sfrac{\Delta t}{4r_{s}} \right) \right]^{2} \right\} =   \mathcal{O} \left[ \frac{\sigma^2(x,x')}{r_s^2}   \right]  \,,\label{wf6}
\ee
showing that \pref{wf5} is a good approximation so long as $|\sigma(x,x')| \ll r_s^2$, or
\begin{eqnarray}
\left( 1 - \frac{r_{s}}{r_0} \right) \sinh^{2}\left( \frac{\Delta t}{4r_{s}} \right) \ll 1 \ . \label{small}
\end{eqnarray}
Notice that this can remain valid even when $\Delta t / r_{s} \gg 1$ so long as $r_0$ is sufficiently close to $r_s$ that (\ref{small}) remains satisfied. 

\begin{figure}[t]
\centering
\includegraphics[width=0.9\textwidth]{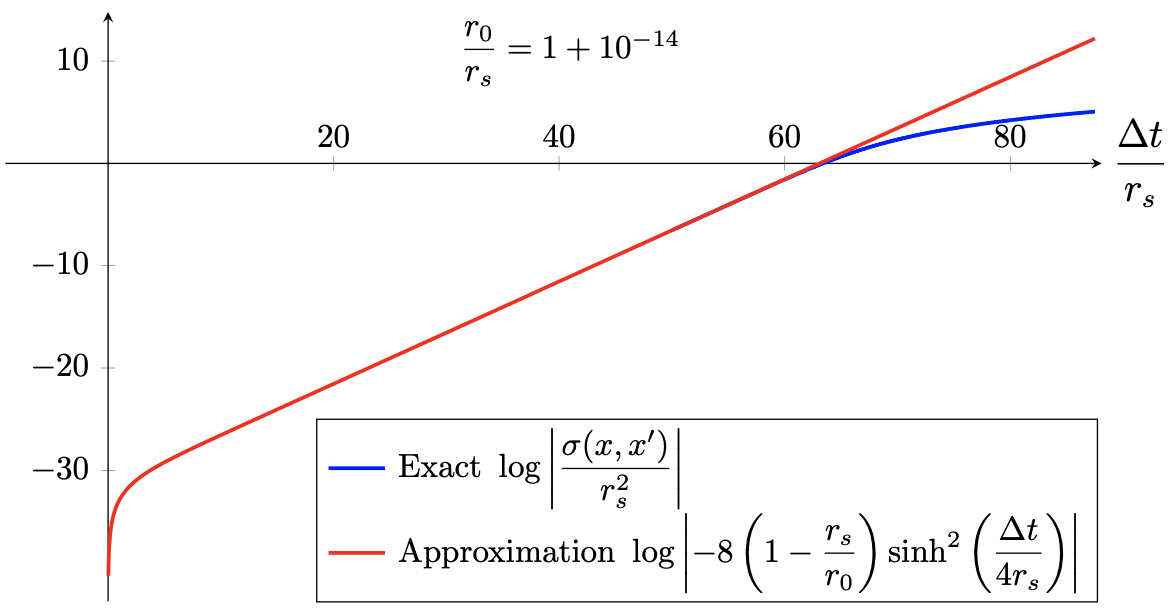}
\caption{\label{fig:Num2} Numerical comparison of the Synge world-function and the asymptotic expressions \pref{wf5} and \pref{wfSchwz}, showing how \pref{wf5} enjoys the broader domain of validity. This plot assumes $r_0/r_s = 1 + 10^{-14}$.}
\end{figure}

Performing the same calculation using Schwarzschild coordinates instead gives 
\be \label{wfSchwz}
   \sigma(x.x') \simeq -\frac{1}{2} \left(1 - \frac{r_s}{r_0} \right) (\Delta t)^2 + \ldots \,.
\ee
Although this agrees with (\ref{wf5}) for $\Delta t  \ll r_s$, the domain of validity of \pref{wf4} turns out to be larger, applying even when  $\Delta t / r_{s}$ is {\it not} small. This can be seen numerically in  $\sigma$, as is shown in Figures \ref{fig:Num1} and \ref{fig:Num2}. Also shown in these figures is how the domain of validity of \pref{wf5} can be extended out to extremely large values of $\Delta t/r_s$ simply by choosing $r_0$ to be ever-closer to the horizon itself.

\end{document}